\documentclass{emulateapj}

\slugcomment{Not to appear in Nonlearned J., 45.}

\shorttitle{A Keplerian gaseous disk around the B0 star R Mon}
\shortauthors{Fuente et al.}

\begin{document}


\title{A Keplerian gaseous disk around the B0 star R Mon}


\author{A. Fuente$^1$, T. Alonso-Albi$^1$, R. Bachiller$^1$, A. Natta$^2$, L. Testi$^2$, R. Neri$^3$, P. Planesas$^1$ }

\altaffiltext{1}{Observatorio Astron\'omico Nacional (OAN), Apdo. 112,
E-28803 Alcal\'a de Henares (Madrid), Spain}
\altaffiltext{2}{INAF-Osservatorio Astrofisico de Arcetri, Largo Enrico Fermi 5, I-50125 Firenze, Italy}
\altaffiltext{3}{Institute de Radioastronomie Millim\'etrique, 300 rue de la Piscine, 38406 St Martin d'Heres Cedex, France}
\email{a.fuente@oan.es}
%
%


\begin{abstract}
We present high-angular resolution observations  
 of the circumstellar disk 
around the massive Herbig Be star R~Mon (M$_*$$\sim$8~M$_\odot$)
in the continuum
at 2.7mm and 1.3mm and the $^{12}$CO 1$\rightarrow$0  
and 2$\rightarrow$1 rotational lines.
Based on the new 1.3mm continuum image
we estimate a disk mass (gas+dust) of 0.007~M$_\odot$ and
an outer radius of $<$150~AU. 
Our CO images  are consistent with the existence of {\it a Keplerian 
rotating gaseous disk around this star}. Up to our knowledge, this is the most 
clear evidence for the existence of Keplerian disks around massive stars reported 
thus far. 
The mass and physical characteristics of this disk are similar to those
of the more evolved T Tauri stars and indicate a shorter
timescale for the evolution and dispersal of circumstellar disks around
massive stars which lose most of their mass before the star becomes visible. 

\end{abstract}





\section{Introduction}

Herbig Ae/Be stars (HAEBE) are intermediate mass (M$\sim$2--10~M$_\odot$) 
pre-main sequence objects. 
A big theoretical and observational effort has been done in recent years for the understanding of 
the disk occurrence and evolution in HAEBE.
For Herbig Ae and late-type Be stars (spectral type later than B7) (hereafter HAE), 
it is generally accepted the presence
of disks similar to those associated with T Tauri stars (TTs). The association of the more
massive early Be stars (HBE) with disks is more uncertain.
Evidence for the existence of dusty and gaseous circumstellar disks around some HBE exists at
optical, NIR and mid-IR wavelengths (Meeus et al. 2001,Vink et al. 2002, Millan-Gabet et al. 2001, 
Acke et al. 2005).
Furthermore, disks around HBE seem to have different geometry from those around
HAE and TTs. While most HAE and TTs have flared disks, 
HBE seem to have (if any) geometrically flat disks (see e.g. Acke et al. 2005).
However, 
the direct detection of circumstellar disks around HBE 
at millimeter wavelengths has remained elusive until recently. 

We carried out a high sensitivity search for circumstellar disks around
HBE with the aim of studying the frequency and time-scales
of disks in these massive stars. We observed a sample of HBE in the continuum 
at mm- and cm- wavelengths using the PdBI and VLA
(Fuente et al. 2001;
Fuente et al. 2003, hereafter Paper~I) and detected two dusty disks 
around MWC~1080 and R Mon with total (gas+dust) masses of
$\sim$0.003~M$_\odot$ and $\sim$0.01~M$_\odot$ respectively.
These were the first detections of dusty disks in HBE.

R~Mon is a very young HBE with a T Tauri companion separated by 0.69$''$.
Because of its youth, R Mon is not directly visible in the optical but appears 
as a resolved conical reflection nebula in scattered light. 
At infrared wavelengths, R Mon appears as a point
source located 0.06$"$$\pm$0.02 south from the optical peak. 
Close et al. (1997) measured 
an extinction of A$_v$=13.1 mag 
towards the star that they interpreted as due to an optically thick disk 
of radius, r$_{out}$=100~AU.
Our previous VLA and PdBI images of R~Mon  at 1.3cm, 0.7cm and 2.7mm showed
a compact source located $\sim$0.45$"$ south from the optical position (Paper~I).
Based on these interferometric observations we derived a mass of 0.01~M$_\odot$
for the dusty disk. R Mon hosts the most massive disk detected in HBE so far
and it is the best candidate to study disks around massive stars.
In this paper, we present observations at higher angular resolution 
and sensitivity of the R~Mon disk
in the continuum at 3mm and 1.3mm and in the $^{12}$CO rotational lines.
Our $^{12}$CO observations are consistent with the existence of
a gaseous disk in Keplerian rotation around the star. 

\section{Observations}

We present high-angular resolution observations in the continuum
at 115.3 GHz and 230.5 GHz and in the $^{12}$CO 1$\rightarrow$0 and
2$\rightarrow$1 lines towards the B0 star R~Mon.
The $^{12}$CO observations were carried out with the IRAM\footnote{IRAM is 
supported by INSU/CNRS (France), MPG (Germany) and IGN (Spain). } array at 
Plateau de Bure (PdBI)  from 2003, December to 
2004, March in the C and A configuration providing a beam 
of 2.28$"$$\times$1.27$"$ at 115.3 GHz and  
1.47$"$$\times$0.94$"$ at 230.5 GHz. 
The absolute flux density scale was determined 
from measurements on MWC349, 0420-014 and 0528+134. 
The maps were not corrected for primary beam attenuation.
During 2006, January we carried out continuum observations 
of this source using the new A$^+$ configuration of PdBI. These observations
provide a beam  of 1.28$"$$\times$0.83$"$ at 2.7mm and
0.72$"$$\times$0.33$"$ at 1.3mm. 
All the images are centered at 
R.A.= 06:39:09.95, Dec=+08:44:09.6 (J2000).
	
\setlength\unitlength{1cm}
\begin{figure}
\includegraphics{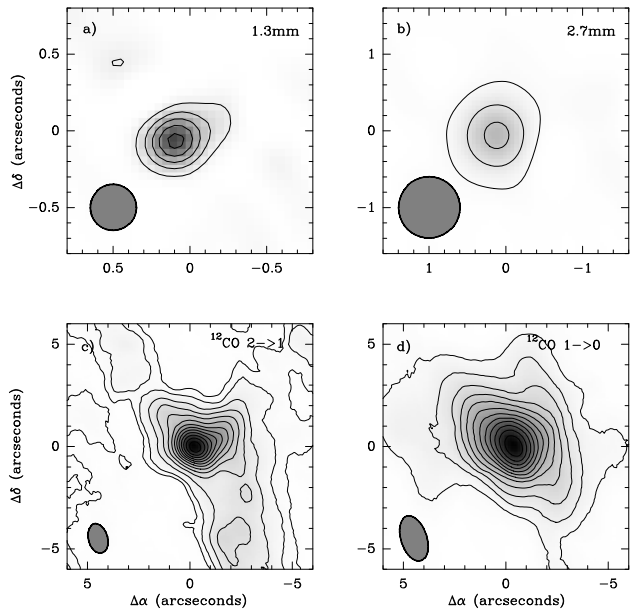}
\vspace{10cm}
\caption{Fig 1ab show the
interferometric images of the circumstellar disk around R~Mon
in the continuum at 1.3mm (a) and 2.7mm (b). Contour levels
are 1.5mJy/beam to 8mJy/beam by 1.5mJy/beam in 
the 1.3mm image and
0.75mJy/beam to 8mJy/beam by 1.5mJy/beam in the 3mm image. The
images have been obtained by convolving the clean components
with a 0.3$"$ circular beam in the case of the 1.3mm map and
a  0.8$"$ circular beam in the case of the 2.7mm image.
In Fig. 1cd we show
integrated  intensity maps of the $^{12}$CO 2$\rightarrow$1 (c)
and $^{12}$CO 1$\rightarrow$0 (d) rotational lines.  
The contour levels  are 0.25 , 0.5  to 6  by steps of 0.5  Jy/beam~km/s
in the $^{12}$CO 2$\rightarrow$1 image and
0.25  to 6 by 0.25 Jy/beam~km/s in the
$^{12}$CO 1$\rightarrow$0 image. The beam is drawn in the
bottom left corner of each panel.
 }
\end{figure}

\setlength\unitlength{1cm}
\begin{figure}
\includegraphics{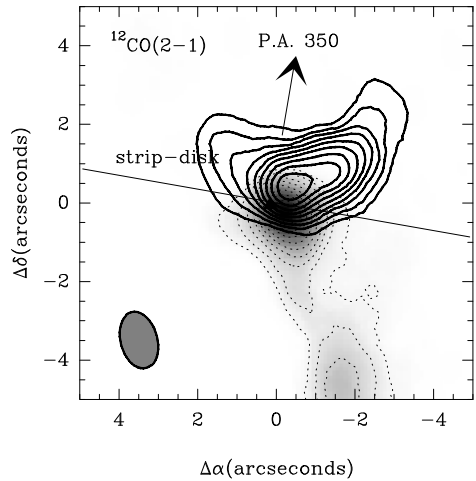}
\vspace{8cm}
\caption{ Thick
contours show the
integrated intensity emission of the  $^{12}$CO 2$\rightarrow$1 rotational
line in the velocity interval 3.6--7~km/s and  dashed contours and grey scale map, the 
integrated intensity emission in the velocity interval 12-15.4~km/s. 
The arrow indicates the direction of the outflow axis.
The studied strip  is drawn in
the Figure. }
\end{figure}

\section{Continuum images}
The new high angular resolution  continuum images are presented in Fig.~1ab. 
These images provide a more accurate position for the disk
that now is settled to R.A.=06:39:09.954  Dec=+08:44:09.55 (J2000).  
The peak emission in the new 2.7mm continuum image 
is 4.1$\pm$0.5~mJy/beam
in agreement with our previous measurement
published in Paper~I. We have
detected, for the first time, the R~Mon disk at 1.3mm.
The peak flux in the 1.3mm image is 7.2$\pm$1.0 mJy/beam and the
integrated flux is 11.8 mJy.
This difference between the peak flux and the integrated
flux at 1.3mm suggests a spatial extension  
of $\sim$0.3$"$. At the distance of R~Mon, this implies a 
disk radius of $\sim$150~AU. Our new 1.3mm measurement
is consistent with our previous upper limit 
already discussed in Paper I and implies a
spectral index for the dust of $\beta$$<$0.5.
The new value of the 1.3mm flux can be used to further
constrain the disk mass. 
After subtracting the free-free emission, we have
fitted the emission at millimeter wavelengths
assuming thermal dust emission
with T$_d$=215~K (Natta et al. 2000), 
$\kappa = 0.01 ( \lambda (mm) / 1.3 )^{-\beta} cm^{-2} g^{-1}$
and a gas-to-dust ratio of 100 (see more details in Paper I).   
Our best fit is obtained for a disk with a mass of 0.007~M$_\odot$ 
and $\beta$=0.3--0.5. 
Values of $\beta$ between 0.5 and 1
are usually found in circumstellar
disks around HAE and TTs and are thought to be
evidence for grain growth in these disks (see e.g.
Natta et al. 2006).
The low value of $\beta$ in R~Mon suggests that
grain growth has proceeded to very large sizes already in the short lifetime
of its disk.

\section{CO images}
We have carried out 
high angular resolution mapping of the 
$^{12}$CO J=1$\rightarrow$0 and
J=2$\rightarrow$1 lines towards R~Mon using
the PdBI. Since the interferometer does
not sample low spatial frequencies,  almost all the
flux is missed  at the cloud velocities
(between 7.0 and 12.0~km~s$^{-1}$).
However, intense molecular emission is detected at the
velocity intervals  of 3.6~km~s$^{-1}$$<$v$<$7.0~km~s$^{-1}$
and   12.0~km~s$^{-1}$$<$v$<$15.6~km~s$^{-1}$ (see Fig. 3).
These $^{12}$CO observations allow us to study in detail
the gas kinematics around R~Mon.

\subsection{ Extended component: The red filament}
In Fig 1cd we present the integrated intensity maps 
of the $^{12}$CO rotational
lines. The maps show a compact molecular clump 
and a more extended weak component. 
The size of the compact clump in 
the map of the $^{12}$CO 2$\rightarrow$1 line 
is $\sim$2.5$"$ (2000~AU at a distance of 800~pc).
The weak component is extended in the North-South (N-S) 
direction. Towards the North,
its emission presents a parabolic shape following
the border of the infrared reflection
nebula. Towards the South, the emission
is concentrated in an elongated feature. 
Since the emission at the velocity of the molecular cloud has
been filtered out by the interferometer, our integrated intensity maps
are actually the sum of the emission of the blue-shifted and 
red-shifted gas.  

In Fig.~2 we show the integrated intensity maps of
the emission at blue-shifted and red-shifted velocities separately. 
The emission 
presents a clear bipolar morphology with the blue
lobe towards the North and the red one towards the South. This
N-S distribution is consistent with the morphology of the bipolar outflow at
large scale (Cant\'o et al. 1981, Bachiller et al. 1987). 
The red lobe presents a jet-like morphology
and remains unresolved in the direction perpendicular to it.
Jets in radio continuum at 1.3 cm (Paper I) and [SII] emission 
(Movsessian et al. 2002) have  previously been detected in R~Mon.
However, the direction of the red filament does not coincide
with the direction of any of these jets but with the direction of one 
of the walls of the cavity suggesting that this high velocity
emission arises in the entrained gas
of the molecular cloud that is being accelerated by the outflow.
 
\begin{table}
\caption{Model results$^*$}
\begin{tabular}{lcc}\\  \hline \hline
\multicolumn{1}{c}{} &  \multicolumn{1}{c}{Flat disk} &
\multicolumn{1}{c}{Flared disk} \\ \hline     
Stellar Mass (M$_\odot$)    &  8$\pm$1                   &     8$\pm$1  \\
i($^\circ$)                             &  20$\pm$5                 &     20$\pm$5  \\
$r_0$  (AU)                          &      1                           &      0.8              \\                        
$T_0$ (K) (r=r$_0$)             &   4500$\pm$500       &      3200$\pm$500 \\
$q$                                       &   0.62$\pm$0.03      &     0.66$\pm$0.02  \\
$M_d$ (M$_\odot$)              &  0.014$\pm$0.001  &      0.08$\pm$0.01  \\
$p$                                       &    1.3$\pm$0.1         &  0.8$\pm$0.1 \\
H(r)                                       &                                 &  0.33+0.83 r (AU) \\
$\Delta v_{turb}$  (km/s)     &   0.8$\pm$0.2           & 0.0$\pm$0.3  \\
$\chi^2$$^{\$}$                                &  1.5                          &   1.8  \\
\hline \hline
\end{tabular}\\
$^*$Errors are those obtained from the fitting procedure and do not account
for other uncertainties inherent in this kind of calculations (instrumental
errors, optical depth effects...).\\ 
$^{\$}$$\chi^2=\frac{1}{N} \sum_{N,v} \frac{(T_b(model)-T_b(obs))^2}{\sigma^2}$
where $N$ is the number of positions and $\sigma$, the rms of
each spectrum.
\end{table}
\subsection{Compact component: A Keplerian gaseous disk}

A strong
compact clump is detected towards the star
in the CO images. Our interferometric 
CO observations allow for the first time to study
the kinematics of the molecular gas at the scale of
the circumstellar disk. In Fig.~3 
we show the Position-Velocity (P-V) diagrams of the emission of the
$^{12}$CO 1$\rightarrow$0 and
2$\rightarrow$1  rotational lines in a strip perpendicular to
the outflow (see Fig.~2). 
The P-V diagrams show the characteristic 
$``$butterfly" shape of a rotating
disk. 

We have modeled 
the emission of the $^{12}$CO 1$\rightarrow$0 and
2$\rightarrow$1 lines to have a deeper insight into
the kinematics and physical characteristics of the disk.
Our model assumes Local Thermodynamic Equilibrium
(LTE), a standard $^{12}$CO abundance X(CO)=8~10$^{-5}$
and radial power laws for the gas kinetic temperature and the 
surface density distribution  ($T_{k}=T_0~(r/r_0)^{-q}$ and  
$\Sigma = \Sigma_0 ~(r/r_0)^{-p}$).

Since the CO emission has been filtered
at the cloud velocities, we have only fitted  the high velocity 
emission (3.6 km s$^{-1}$$<$v$<$7.0 km s$^{-1}$
and 12.0 km s$^{-1}$$<$v$<$15.6 km s$^{-1}$).
Based on NIR and mid-IR observations, several authors
have proposed that contrary to HAE and TTs that host flared disks, 
HBE seem to have geometrically flat disks (see e.g. Acke et al. 2005).
We have considered the two different geometries, a flared
and a flat disk.  The best fit with each geometry is shown in 
Table 1. 

The intensity ratio between the $^{12}$CO 1$\rightarrow$0 
and 2$\rightarrow$1 lines is $\sim$1 as expected for  
optically thick emission.
Since the $^{12}$CO lines are
optically thick, the $` `$butterfly" shape is only
dependent on the gas kinematics 
regardless of the disk geometry. 
The observational data are well fitted with 
a disk in Keplerian rotation
around the star. The outer radius of the disk
is 1500~AU, the inclination angle 20$^\circ$$\pm$5 
and the mass of the star is constrained to 8$\pm$1~M$_\odot$. 
The derived spectral types for R~Mon range from B0 
(see e.g. Hillenbrand 1992) to B8 
(Mora et al. 2001). The stellar mass we obtain is 
consistent with a star of a spectral type between
B0 and B3 but it is larger than that expected for 
a B8 star. The high gas kinetic temperature derived
using both the flat and the flared geometry
also supports the
classification of R~Mon as an early Be star (see Table 1)
rather than as a cooler B8 object.

Since the $^{12}$CO lines are
optically thick, the mass of the
gaseous disk is not well determined by our observations. 
In fact,  the estimated value  of the mass is 
strongly dependent 
on the assumed geometry.  Using a flat disk, the best fit  
is obtained with 
M$_d$=0.014~M$_\odot$. 
A different disk mass is obtained with the flared geometry. 
In our flared model the  height (H) increases linearly
with the radius  (H~$\propto$~$r$) and the
density is assumed to be constant in the 
z-direction. These assumptions are reasonable since in
hydrostatic equilibrium H~$\propto$~$r^{9/8}$. 
Moreover, Pi\'etu et al. (2003)  based on interferometric
CO observations obtained that 
H~$\propto$~$r^{1.2}$  in the disk around the Ae star HD~34282.
We obtain a reasonable fit to the data with H varying from
1 AU (at r=0.8 AU) to 1250 AU (at r=1500 AU),
and M=0.08 M$_\odot$.
This mass is 10 times larger than that obtained from 
dust emission and a factor of 6 larger than
that obtained from the CO lines assuming a flat disk.
This is quite surprising since in TTs and HAE 
the mass derived from the CO lines
(assuming a standard CO abundance)
is smaller than that derived from continuum
observations (see e.g. Pi\'etu et al. 2003) and suggests 
that the disk associated with R~Mon is very
likely flat or  at least flatter than those associated
with TTs and HAE. This conclusion is also supported by the low
value of $\chi^2$ obtained in the case of the flat
disk. 
However, we cannot discard the 
possibility of a flared disk since 
the dust emissivity in the disk around R~Mon
could be smaller than in those around TTs and
HAE because of the rapid grain growth and the
fit we obtain using the flared disk is still reasonably good.

The values of T$_0$, $q$ , $p$ and $\Delta v_{turb}$
are similar in the flat  and the flared disk
models. 
Moreover, the values of $p$, $q$ and
$\Delta v_{turb}$ are similar to those measured in
the disks associated with TTs and HAE (see e.g. Dutrey et al. 2004)
suggesting that all these stars host disks with 
similar physical characteristics.

In Fig.~3 and 4 we compare the results
of our flat disk model with the observations. 
The model fits
very well the observations in the $``$red$"$ part of the P-V diagram.  
However there is
a displacement between the observed emission 
and that predicted by the model in the $``$blue$"$ part (see Fig.~3 right).
The only way to fit the blue part would be to assume an asymmetrical disk
and/or an off-center star.

\setlength\unitlength{1cm}
\begin{figure}
\includegraphics{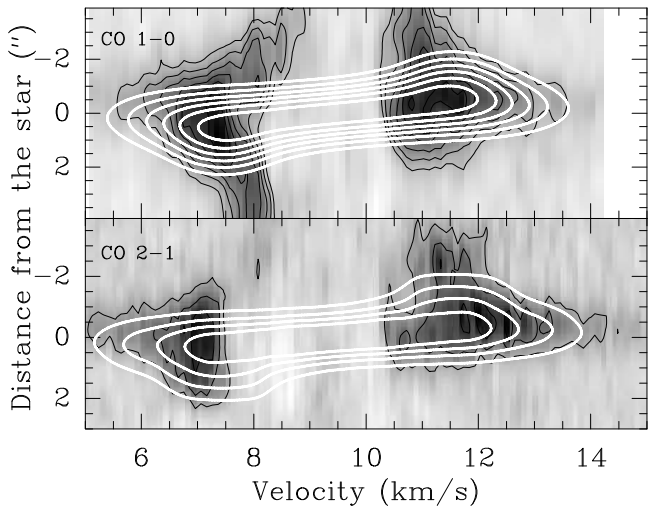}
\vspace{8cm}
\caption{ 
Grey scale images are the Position-Velocity (P-V) diagrams of the CO rotational
lines along the strip drawn in Fig.~ 2.  White contours show  
the synthesized P-V diagram with our flat disk model. Contour levels are
200 mJy/beam to 800 mJy/beam in steps of 100 mJy/beam (3.16~K) for the CO 1$\rightarrow$0
diagram and 200 mJy/beam to 800 mJy/beam in steps of 200 mJy/beam (3.46~K)
for the CO 2$\rightarrow$1 one. }
\end{figure}

\setlength\unitlength{1cm}
\begin{figure}
\includegraphics{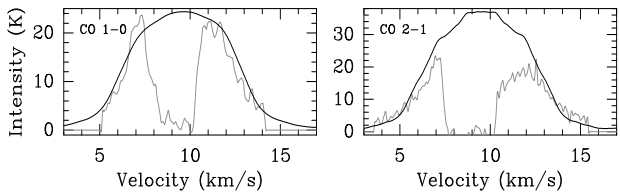}
\vspace{3.0cm}
\caption{ 
Comparison between the spectra observed towards the star position (grey line)
and those predicted by our flat disk model (black line). 
Note that the interferometer misses the flux at the cloud velocities. 
}
\end{figure}

\section{Discussion and Conclusions}

R~Mon is the first disk around a HBE detected in
molecular lines and so far, our unique
opportunity to investigate the physical structure 
and kinematics of the gas reservoir in the disks
associated with these stars. Our most interesting result
is that the disk is in Keplerian rotation
around the star. Keplerian rotation has also been 
found in most of the TTs and HAE studied so far
and indicates a similar formation mechanisms for
the stars in the range 1--8 M$_\odot$. 
The values obtained for the indexes $q$ and $p$  
as well as that for the  turbulent broadening are also similar 
to those found in TTs and 
some HAE (see e.g. Dutrey et al. 2004). R~Mon
is a very young object (age$\sim$10$^5$ yr) which is
still deeply embedded in the molecular cloud. 
On the contrary, TTs are evolved objects  
(age$>$10$^6$ yr) which
are not any more associated with the parent
molecular core (age$>$10$^6$ yr). The
similarity between the disks of such different
objects suggests a shorter timescale for the evolution
of the disks associated with HBE. 
The low occurrence of disks associated with HBE also supports 
this conclusion. 
In Paper~I we showed that the disk masses in HBe stars are at least an order
of magnitude lower than in HAe stars. In fact, the ratio M$_d$/M$_*$ is 
roughly constant and equal to 0.04 for
stars with spectral type A0-M7 and M$_d$/M$_*$ $<$ 0.001 in
HBe stars. This indicates a rapid dispersal
of the disk material in HBE.    

We have compared the R~Mon disk with those associated with
massive protostars in order to have a deeper insight into the
evolution of disks around HBE.
Evidences for circumstellar disks have been found in massive
protostars. The surface density
in these disks is well described with $p>$2 
(Cesaroni et al. 2005). This means that most 
of the gas is concentrated 
at small radii. 
Recently, Schreyer et al. (2006) detected a 1 M$_\odot$ disk around 
MWC~490,  a 8--10 M$_\odot$ star which is in a
transition stage to HBE. In this case, the surface density
varies with p$=$1.5 and the disk radius is 1400 AU. The
values of $p$ and $r_{out}$ are similar to those 
derived in the disk around R~Mon. 
However the mass of the MWC~490 disk is 
more than an order of magnitude larger
than that of the R~Mon disk. These results
indicate that the disks around HBE
flatten and lose a large fraction of their mass
($\sim$90\% )  before the pre-main sequence phase ($<$10$^5$ yr).
The flattening of the disks 
can be due to the rapid grain growth.
The low value of $\beta$ measured in R~Mon is a clear evidence
of the presence of large grains. 
The grain growth causes the optical depth of the
disk to drop and allows the UV radiation to penetrate
deep into the circumstellar disk and photo-evaporate the disk external
layers (Dullemond \& Dominik 2004).  Thus, we can propose
an evolutionary sequence in which
the disks associated with HBE
start with a flaring shape but become
flat during the pre-main sequence
and lose most of their mass ($>$90\%) before the star
becomes visible ($<$10$^5$ yr).
The results of this Letter are consistent with this scenario  
which, however,
needs to be confirmed.

\acknowledgements
This work has been partially supported by the Spanish MEC and 
Feder funds under grant ESP2003-04957 and by SEPCT/MEC 
under grants AYA2003-07584 and AYA2002-01055.

\end{document}